\documentclass{elsart}

\usepackage{natbib}

\usepackage{amssymb}

\begin{document}

\begin{frontmatter}

\title{Feature selection for high dimensional data\\ in astronomy}

\author{Hongwen Zheng$^a$ and Yanxia Zhang$^b$}

 \address[label1]{Institute of Mathematics and Physics, North China
Electric Power University, Deshengmenwai, Zhuxinzhuang, Beijing,
102206, China}
\address[label2]{National Astronomical
Observatories, CAS, 20A Datun Road, Chaoyang District, Bejing
100012 China}

\begin{abstract}
With an exponentially increasing amount of astronomical data, the
complexity and dimension of astronomical data are likewise growing
rapidly. Extracting information from such data becomes a critical
and challenging problem. For example, some algorithms can only be
employed in the low-dimensional spaces, so feature selection and
feature extraction become important topics. Here we describe the
difference between feature selection and feature extraction methods,
and introduce the taxonomy of feature selection methods as well as
the characteristics of each method. We present a case study
comparing the performance and computational cost of different
feature selection methods. For the filter method, ReliefF and fisher
filter are adopted; for the wrapper method, improved CHAID, linear
discriminant analysis (LDA), Naive Bayes (NB) and C4.5 are taken as
learners. Applied on the sample, the result indicates that from the
viewpoints of computational cost the filter method is superior to
the wrapper method. Moreover, different learning algorithms combined
with appropriate feature selection methods may arrive at better
performance.

\end{abstract}


\end{frontmatter}

\section{Introduction}

Driven by the enormous technological advances in telescopes and
detectors, the exponential increase in computing capabilities, and
the fundamental changes in the observing strategies used to gather
data, astronomy is undergoing a revolutionary shift, and entering a
data flood and information-rich era. The volumes of astronomical
data amount to many Terabytes, even Petabytes, from which catalogs
or images of many millions, or even billions of objects are
extracted. For each object, some tens or even hundreds of parameters
are measured. With the Global Virtual Observatory (GVO) coming into
implementation step by step, the science based on Virtual
Observatory (VO) may be done in the image domain, and also the
interaction between the image and catalog domains. What is more
important is that, much of the science will be done purely in the
catalog domain of individual or federated sky surveys. A typical
data set may be a catalog of $\sim 10^8-10^9$ sources with
$\sim10^2$ measured attributes each, i.e., a set of $\sim 10^9$ data
vectors in a $\sim 100$-dimensional parameter space. Moreover, only
the dimension of spectral data adds up to many thousands or even
larger. Astronomy may become such a data-rich subject as other
subjects, e.g. biology.

The recent increase of dimensionality of data poses a severe
challenge to many existing data mining, pattern recognition, machine
learning, artificial intelligence methods as well as feature
selection/extraction methods with respect to efficiency and
effectiveness. The problem is especially severe when large
databases, with many features, are searched for patterns without
filtering of important features based on prior knowledge. The
growing importance of knowledge discovery and data mining methods in
practical applications has made the feature selection/extraction
problem a quite hot issue, especially when mining knowledge from
databases or warehouses with huge amounts of records and columns.
Feature selection/extraction, as a preprocessing step to data
mining, image processing, conceptual learning, machine learning,
etc, has been effective in reducing dimensionality, removing
irrelevant and redundant data, increasing learning accuracy, and
improving comprehensibility. Based on these merits, it is an
important and necessary preprocessing step before the implementation
of algorithms. So far feature selection/extraction has played
important roles in many data mining tasks, such as classification
(Dash \& Liu, 1997), clustering (Dash et al. 2002) and regression
(Miller 2002). A lot of work has been carried out on feature
selection/extraction in astronomy. For instance, Re Fiorentin et al.
(2007) used principal component analysis (PCA) on pre-processing of
star spectra, then estimated stellar atmospheric parameters.
Ferreras et~al. (2006) employed PCA to the star formation history of
elliptical galaxies in compact groups. Lu et al. (2006) put forward
ensemble learning for independent component analysis (EL-ICA) on the
synthetic galaxy spectra. EL-ICA sufficiently compressed the
synthetic galaxy spectral library to six nonnegative independent
components (ICs), which are good templates for modeling huge amounts
of normal galaxy spectra. Zhang et al. (2004) implemented ReliefF
algorithms for feature selection and then found that the naive Bayes
classifier based on ReliefF algorithms is robust and efficient to
preselect AGN candidates. Zhang \& Zhao (2004) used histogram as
feature selection technique to evaluate the significance of the
considered features for classification.

\section{Feature selection and feature extraction}

For data mining methods that only execute in the low-dimensional
spaces, feature selection or feature extraction is a necessary step
before they can deal with high dimensional data. Feature selection
is concerned with locating a minimum subset of the original features
that optimizes one or more criteria, rather than producing an
entirely new set of dimensions for the data. Feature extraction
(i.e. feature transformation) is a preprocessing technique that
transforms the original features of a data set to a smaller, more
compact feature set, while retaining as much information as
possible. Usually, feature selection approaches are divided into
three types: filter, wrapper and embedded methods; feature
extraction approaches include principal component analysis (PCA),
linear discriminant analysis (LDA), independent component analysis
(ICA), latent semantic index (LSI) and so on. Often, feature
extraction precedes feature selection; first features are extracted
from the data and then, some of the extracted features with low
discriminatory power are discarded, leading to the selection of the
remaining features. Notice that the two techniques are also
complementary in their goals; feature selection leads to savings in
measurement cost and the selected features retain their original
physical interpretation. On the other hand, the transformed features
obtained by feature extraction techniques may provide a better
discriminatory ability than the best selected subset, but these
features fail in retaining the original physical interpretation and
may not have a clear meaning.

\section{Taxonomy of feature selection methods}

In order to evaluate the selected subset, the characteristics of the
data, the target concept and the learning algorithm should be taken
into account. Based on these information, the methods of feature
selection can be classified into three categories: filter methods,
wrapper methods and embedded methods. For good reviews about
existing methods for feature selection, readers can refer to Liu \&
Motoda (1998), Guyon \& Elisseeff (2003).

Filter methods are simplest and most frequently used in the
literature. They consist of feature ranking algorithms (e.g. Relief
presented by Kira \& Rendell in 1992) and subset search algorithms
(e.g. Focus given by Almuallim \& Dietterich in 1994). For filter
methods, features are scored according to the evidence of predictive
power and then are ranked. The top $s$ features with the highest
scores are selected and used by the classifier. The scores can be
measured by t-statistics, F-statistics, signal-noise ratio, etc. The
number of features selected, $s$ , is then determined by cross
validation. Advantages of filter methods are that they are fast and
easy to interpret. The characteristics of filter methods are as
follows:

(1): Features are considered independently.

(2): Redundant features may be included.

(3): Some features which as a group have strong discriminatory power
but are weak as individual features will be ignored.

(4): The filtering procedure is independent of the classifying
 method.

Wrapper methods use iterative search. Many ``feature subsets" are
scored based on classification performance and the best is used. The
approaches of subset selection contain forward selection, backward
selection, their combinations. The problem is very similar to
variable selection in regression. For example, exhaustive searching
is impossible; greedy algorithms are used instead; confounding
problem can happen in both scenarios. Exhaustive search finds a
solution by trying every possibility. A greedy algorithm might also
be called a ``single-minded" algorithm or an algorithm that consumes
all of its favorites first. The idea behind a greedy algorithm is to
perform a single procedure in the recipe over and over again until
it can't be done any more and see what kind of results it will
produce. It may not completely solve the problem, or, if it produces
a solution, it may not be the very best one, but it is one way of
approaching the problem and sometimes yields very good (or even the
best possible) results. In regression, it is usually recommended not
to include highly correlated covariates in analysis to avoid
confounding. But it's impossible to avoid confounding in feature
selection of microarray classification. A detailed overview of
wrapper methods is introduced by Kohavi \& John (1997). The
characteristics of wrapper methods are listed below:

 (1): Computationally expensive for each feature subset considered,
the classifier is built and evaluated.

 (2): As exhaustive searching is impossible, only greedy search is applied. The advantage of greedy search is simple and quickly to
find solutions, but its disadvantage is not optimal, and susceptible
to false starts.

 (3): It is often easy to overfit in these methods.

Finally another type of feature subset selection is identified as
embedded methods. In this case, the feature selection process is
done inside the induction algorithm itself, i.e. attempting to
jointly or simultaneously train both a classifier and a feature
subset. They often optimize an objective function that jointly
rewards the accuracy of classification and penalizes the use of more
features. Intuitively appealing examples are nearest shrunken
centroids, CART and other tree-based algorithms. Common practice of
feature selection is to use the whole data, then apply
cross-validation (CV) only for model building and classification.
However, usually features are unknown and the intended inference
includes feature selection. Then, CV estimates as above tend to have
a downward bias. Feature selection should be done only from the
training set used to build the model (and not the entire set).

Embedded methods are done within the learning algorithm preferring
some features instead of others and possibly not including all the
available features in the final model induced by the learning
algorithm. However, filter and wrapper methods are located one
abstraction level above the embedded one, performing a feature
selection process for the final model apart from the embedded
selection done by the learning algorithm itself.

Another category of approaches called feature weighting
approaches, is not always considered in the classical
classification of feature selection methods. In the implementation
process of these methods, actual feature selection is substituted
by a feature weighing procedure able to weight the relevance of
the features.

In brief, application of the filter method requires computational
complexity, but the higher complexity of the wrapper method will
also produce higher accuracy in the final result. The filtering
method is a very flexible one, since any target learning algorithm
can be used in conjunction, while the wrapper method is strictly
dependent on the learning algorithm; the filter method is faster,
the selection process is better from the computational point of
view. Embedded approaches are intrinsic to some learning algorithm
and so only an algorithm design with this characteristic can be
used. Finally, if a weighting scheme can be devised, feature
selection can be implemented via feature weighting, by postponing
the selection as a subsequent possible choice using the weights.

\section{Case study}

The data is adopted from Zhang \& Zhao (2004), including 1,656
active galactic nuclei (AGNs), 3,718 stars and 173 normal galaxies.
In this investigation, the plausibility is based on the optical
classification, X-ray characteristics such as hardness ratios and
the extent parameter, and the infrared classification. In order to
classify sources, we consider data from optical, X-ray, and infrared
bands. The chosen parameters from different bands are $B-R$ (optical
index), $B+2.5logCR$ (optical-X-ray index), $CR$ (source countrate),
$HR1$ (hardness ratio 1), $HR2$ (hardness ratio 2), $ext$ (source
extent), $extl$ (likelihood of source extent), $J-H$ (infrared
index), $H-K{\rm s}$ (infrared index), and $J+2.5logCR$
(infrared-X-ray index). Based on these parameters, we may study the
clustering properties of astronomical objects in a multidimensional
parameter space and discriminate AGNs from stars and normal
galaxies. With known samples to construct classifiers by automated
methods, we will effectively preselect source candidates for large
survey projects.

We mainly compare the filter method and the wrapper method in this
section. Feature selection is carried out to study the effect on the
performance of a range of classification algorithms with the
selected attributes. When applying the wrapper method for feature
selection, we used 10-fold cross-validation (CV). While for
classification, two thirds of the sample (3,328) are for training,
one third (2,219) for testing. Fisher filter and ReliefF are used as
filter methods for feature selection. Fisher filter uses an ANOVA
(analysis of variance) for predictive attribute evaluation. A key
idea of the original Relief algorithm (Kira and Rendell, 1992) is to
estimate the quality of attributes according to how well their
values distinguish between instances that are near to each other.
The Relief algorithm assigns high scores to features that match on
near hits and don't match on near misses (in the context of nearest
neighbor classification) (Robnik-$\check{S}$ikonja \& Kononenko
2003). Improved CHAID, linear discriminant analysis (LDA), Naive
Bayes (NB) and C4.5 are taken as learners in order to do this case
study. CHAID (chi-squared automatic interaction detection,
Rakotomalala \& Zighed 1997) belongs to decision tree family,
applies $\chi^2$ test during decision process, its main
characteristics is forward-pruning and multiple-branch. LDA (Saporta
1990) maximizes the ratio of between-class variance to the
within-class variance in any particular data set thereby
guaranteeing maximal separability. LDA doesn't change the location
but only tries to provide more class separability and draw a
decision region between the given classes. This method also helps to
better understand the distribution of the feature data. The naive
Bayes classifiers assign the most likely class to a given example
described by its feature vector (Mitchell 1997; Zhang et al. 2004).
The classifiers assume that the effect of an variable value on a
given class is independent of the values of other variable. This
assumption is called class conditional independence. It is made to
simplify the computation and in this sense considered to be
``naive". C4.5 is a software extension of the basic ID3 algorithm
designed by Quinlan (1993), and solves issues that are not addressed
by ID3, e.g. C4.5 can handle missing value and continuous features.

\subsection{Selected features}

To be short, we name the attributes: $B+2.5logCR$, $J+2.5logCR$,
$B-R$, $HR2$, $H-K_{\rm s}$, $ext$, $J-H$, $logCR$, $HR1$, $extl$ as
A1, A2, A3, A4, A5, A6, A7, A8, A9, A10, respectively. The
attributes are selected by different feature selection methods, as
shown in Table 1. The attributes marked by symbol ``tick" are
important features identified by different feature selection
methods. Table 1 shows that different features are selected and the
number of features is reduced for different feature selection
methods, both the filter method and the wrapper method.

\begin{table*}[ht]
\begin{center}
\caption{Selected features resulting from different feature
selection methods}
\bigskip
\begin{tabular}{|c|c|c|c|c|c|c|c|c|c|c|}
\hline
 Methods &A1&A2&A3&A4&A5&A6&A7&A8&A9&A10\\
\hline
 ReliefF &$\surd$&$\surd$&$\surd$&$\surd$&$\surd$&$\surd$&&&&\\\hline
 fisher filter&$\surd$&$\surd$&$\surd$&$\surd$&$\surd$&$\surd$&&$\surd$&&$\surd$\\\hline
 improved CHAID &$\surd$&$\surd$&$\surd$&&$\surd$&&&$\surd$&&\\\hline
 LDA &$\surd$&$\surd$&$\surd$&$\surd$&$\surd$&&$\surd$&$\surd$&$\surd$&$\surd$\\\hline
 NB &$\surd$&$\surd$&$\surd$&$\surd$&$\surd$&$\surd$&&&$\surd$&\\\hline
 C4.5& $\surd$&$\surd$&&&$\surd$&&&$\surd$&$\surd$&\\
 \hline
 \end{tabular}
\bigskip
\end{center}
\end{table*}

\subsection{Computational time}

The configuration of the computer used is Microsoft Windows XP,
Pentium (R) 4, 3.2~GHz CPU, 1.00~GB memory. The time to select
features by different methods is indicated in Table 2. For filter
methods (i.e. ReliefF and fisher filter), times required for feature
selection are 26.34~s and 16~ms, respectively. For wrapper methods
using improved CHAID, LDA, NB and C4.5 as learners, times required
are 294.39~s, 31.83~s, 165.27~s and 284.33~s, respectively. Of these
methods, fisher filter spends the least time for feature selection,
only 16~ms, whereas, improved CHAID and C4.5 spend the most time,
more than 280~s. Thus, the speed of the filter method for feature
selection is faster than that of the wrapper method.

Although both ReliefF and fisher filter are filters, they spend
different time to fulfill the task. This is mainly due to the
different principals of the two methods. Fisher filter employs an
ANOVA for feature selection, while Relief algorithm needs distance
computation to estimate the quality of attributes. As a result,
fisher filter is very fast. Similarly, there is variable time cost
for the four different wrappers. In terms of speed, LDA is the
fastest of wrappers.

\begin{table*}[ht]
\begin{center}
\caption{Time for feature selection by different feature selection
methods}
\bigskip
\begin{tabular}{|c|c|}
\hline
Methods  &Time for feature selection (second)\\
\hline
  ReliefF&26.34\\\hline
  fisher filtering&0.016\\\hline
  improved CHAID&294.39\\\hline
  LDA&31.83\\\hline
  NB&165.27\\\hline
  C4.5&284.33\\
   \hline
 \end{tabular}
\bigskip
\end{center}
\end{table*}

\subsection{Accuracy}

We carry out a systematic study of the effect on the performance of
a range of classification algorithms with the attributes selected
using feature selection methods. The classification algorithms used
are improved CHAID, LDA, NB and C4.5. As shown in Table 3, the
results show that, for the majority of situations, all algorithms
benefit by the selected attributes. The results of feature selection
methods vary with respect to accuracy. The performance of the
wrapper method and ReliefF are illustrative, showing different
behaviors: compared to the accuracy of no feature selection method,
the wrapper method may improve, while RelieF may worsen. For fisher
filter, the accuracy is exceeded except by the learner NB. As for
C4.5, fisher filter appears best; while for LDA and NB, ReliefF
appears best. For improved CHAID, whilst the performance with
wrapper method and fisher filter improves, the performance with
RelieF deteriorates. For NB, while the performance with ReliefF
improves, the performance with other techniques decreases. Whilst
these are not statistically significant, it does indicate that care
must be taken when a pre-processing technique (attribute selection
using feature selection algorithms).

\begin{table*}[ht]
\begin{center}
\caption{The accuracy achieved by three feature selection methods}
\bigskip
\begin{tabular}{|c|c|c|c|c|}
\hline
Methods  &no feature selection&wrapper method&ReliefF&fisher filter\\
\hline
  improved CHAID&97.70\%&97.70\%&97.61\%&98.11\%\\\hline
  LDA&95.36\%&95.45\%&95.85\%&95.58\%\\\hline
  NB&97.66\%&97.52\%&98.15\%&97.61\%\\\hline
  C4.5&97.34\%&97.61\%&97.34\%&97.79\%\\
 \hline
 \end{tabular}
\bigskip
\end{center}
\end{table*}

\section{Conclusion}
Data preprocessing is an important part of effective machine
learning and data mining. Feature selection, as a kind of data
preprocessing, is an effective approach to downsizing data. Feature
selection is a process that chooses an optimal subset of features
according to a certain criterion. There are many merits of feature
selection, such as, to reduce dimensionality and remove noise,
improve learning performance, speed up learning process, improve
predictive accuracy and bring simplicity and comprehensibility of
learned results. In this work, feature selection and feature
extraction are compared and the taxonomy of feature selection
methods is surveyed. Three kinds of methods (i.e. filter, wrapper
and embedded methods) have generally been studied for feature
selection. Filters select subsets of variables as a pre-processing
step, independently of the chosen predictor. Wrappers utilize the
learning machine of interest as a black box to score subsets of
variable according to their predictive power. Embedded methods
perform variable selection in the process of training and are
usually specific to given learning machines. The essential
difference between these approaches is that the last two methods
make use of the algorithms that will be used to build the final
classifiers, while a filter method does not. Moreover, a case study
is presented illustrating the performance of different feature
selection methods. From the result of this case study, the filter
method has lower computational cost compared to the wrapper method
and looks most promising for the ``data avalanche" facing astronomy.
Moreover the filter method of selecting features is independent of
learning algorithms and selected features can be used by any
learning algorithm. This is why much work focuses on developing new
filter methods. Given any learning algorithm, we should choose the
appropriate filter method and its performance can be improved. For
example, in our case, NB and LDA combined with ReliefF is best, C4.5
and improved CHAID combined with fisher filter is best. In general,
filters are computationally less intensive, while wrappers produce
better classifications. Regarding the speed of filter methods and
the accuracy of wrapper methods, hybrid methods have been put
forward in order to take advantage from the aforesaid methods. This
approach represents a new trend in feature selection because it
tries to join the speed of the filter approaches with the accuracy
of the wrapper ones. Feature selection is a rather complex issue. It
is not straightforward to determine which feature selection method
is best. Rather, this depends on the characteristics of data (e.g.
linear or nonlinear distribution, with or without noise, continuous
or discrete features, irrelevant or interrelated attributes), the
number of examples and features, the type of learners, the target
task, and so on. We conclude that the high dimensional problems
faced in astronomy may be easily solved by feature selection
methods. The study of feature selection methods in other fields is
growing rapidly and yielding important results. It is necessary to
bring these to the attention of the astronomical community, so the
result can be applied to its critical problems.

{\bf Acknowledgments} We are very grateful to referees for their
important suggestions and comments that have served to strengthen
this paper. This paper is funded by National Natural Science
Foundation of China under grant No.10473013 and No.90412016.

\end{document}